\include{Commands}
\documentclass[11pt]{article}
\usepackage[utf8]{inputenc}
\usepackage{url,hyperref,lineno,microtype,subcaption}
\usepackage[english]{babel}
\usepackage{psfrag}

\usepackage{amsmath}
\usepackage{amsfonts}
\usepackage{amssymb}
\usepackage{bm}
\usepackage{xcolor}
\usepackage[version=4]{mhchem} 
\usepackage{mathtools}
\usepackage{amsbsy}
\usepackage{graphicx}
\usepackage{caption}
\usepackage{subcaption}
\usepackage[figurename=Figure]{caption}
\usepackage[tablename=Table]{caption}
\usepackage{cleveref}
\usepackage[square,sort,comma,numbers]{natbib}
\usepackage{geometry}
\usepackage{times}
\usepackage[color=green!20]{todonotes}

\usepackage{parskip}
\usepackage{color}
\usepackage{wrapfig}
\renewcommand{\baselinestretch}{1.5}
\usepackage{siunitx}
\usepackage{adjustbox}
\usepackage{tabularx}
\renewcommand{\thetable}{\arabic{table}}
\usepackage{footnote}
\usepackage{threeparttable}
\usepackage{ulem}
\renewcommand{\baselinestretch}{1.5}

\title{\bfseries {Applications and Challenges of Machine Learning to Enable Realistic Cellular Simulations}}
\author{R. Vasan\,$^{1}$, M. P. Rowan\,$^{2}$, C. T. Lee\,$^{1}$, G. R. Johnson\,$^{3}$, P. Rangamani\,$^{1}$,  and M. J. Holst\,$^{4,5,*}$}

\begin{document}

\maketitle
$^{1}$Department of Mechanical and Aerospace Engineering, University of California San Diego, La Jolla, CA 92093, USA\\
$^{2}$Department of Bioengineering, University of California San Diego, La Jolla, CA 92093, USA\\
$^{3}$Allen Institute of Cell Science, Seattle, WA 98109, USA\\
$^{4}$Department of Mathematics, University of California San Diego, La Jolla, CA 92093, USA\\
$^{5}$Department of Physics, University of California San Diego, La Jolla, CA 92093, USA
\\$^{*}$Corresponding Author\\
 $|$Email: mholst@ucsd.edu $|$

\maketitle

\begin{abstract}
In this perspective, we examine three key aspects of an end-to-end pipeline for realistic cellular simulations: reconstruction and segmentation of cellular structures; generation of cellular structures; and mesh generation, simulation, and data analysis.  We highlight some of the relevant prior work in these distinct but overlapping areas, with a particular emphasis on current use of machine learning technologies, as well as on future opportunities.
\end{abstract}

\newpage

\section{Introduction}

Machine learning (ML) approaches, including both traditional and deep learning methods, are revolutionizing biology.
Owing to major advances in experimental and computational methodologies, the amount of data available for training is rapidly increasing.
The timely convergence of data availability, computational capability, and new algorithms is a boon for biophysical modeling of subcellular and cellular scale processes such as biochemical signal transduction and mechanics~\cite{lee_exascale_2018}.
To date, many simulations are performed using idealized geometries that allow for the use of commonly used techniques and software \cite{alimohamadi2018role, vasan2018intracellular, bell2019dendritic, ohadi2019geometric, vasan2019dlite}.
This is historically due to the lack of high-resolution structural data as well as the theoretical and computational challenges for simulations in realistic cellular shapes, due to the complexity of generating high-quality, high-resolution meshes for simulation, and the need to develop specialized fast numerical solvers that can be used with very large unstructured mesh representations of the physical domain.

As biophysical methods have improved, the complexity of our mathematical and computational models is steadily increasing \cite{vasan2019mechanical, rudraraju2016mechanochemical, mihai2017family}.
A major frontier for physics-based study of cellular processes will be to simulate biological processes in realistic cell shapes derived from various structural determination modalities \cite{murphy2016building, lee2019open}.
For biological problems ranging from cognition to cancer, it has long been understood that cell shapes are often correlated with mechanism \cite{rangamani2013decoding, deuling1976red, bartol2015nanoconnectomic, ritz1997synchronous, harris1994dendritic}.
Despite such clear correlations, there remain gaps in our understanding of how cellular ultrastructure contributes to cellular processes and the feedback between cellular structure and function.
Challenges such as the diffraction limit of light and difficulties in manipulation of intracellular ultrastructure constrain the potential scope of what can be achieved experiments.
Much like the partnership between biophysics and molecular dynamics simulations have enabled the modeling of invisible protein motions to shed insights on experimental observations, simulations of cellular processes can also aid in the validation and generation of hypothesis currently inaccessible by experimental methods.
Recently, we and others have shown that, for example, cell shape and localization of proteins can impact cell signaling \cite{bell2019dendritic, rangamani2013decoding, cugno2019geometric, ohadi2019geometric, ohadi2019computational}.

The major bottleneck for the widespread use of cell scale simulations with realistic geometries is not the availability of structural data. 
Indeed, there exist many three-dimensional imaging modalities such as confocal microscopy, multiphoton microscopy, super-resolution fluorescence and electron tomography \cite{huang2016ultra, graf2010imaging}. 
For example, automation of modalities such as Serial Block-Face Scanning Electron Microscopy is already enabling the production of data at rapid rates. 
The bottleneck lies in the fact that much of the data generated from these imaging modalities need to be manually curated before it can be used for physics-based simulations. 
This current \textit{status quo} of manually processing and curating these datasets for simulations is a major obstacle to our progress.
In order to bridge the gap between abundance of cellular ultrastructure data generated by 3D electron microscopy (EM) techniques and simulations in these realistic geometries, innovations in machine learning (ML) methods will be necessary to reduce the time it takes to go from structural datasets to initial models.
There are already many similar efforts at the organ/tissue and connectomics length scales
\cite{lichtman2014big, maher2019accelerating, januszewski2018high}.
In this work, we summarize the main steps necessary to construct simulations with realistic cellular geometries (Fig. 1) and highlight where innovation in ML efforts are needed and will have significant impacts. 
We further discuss some of the challenges and limitations in the existing methods, setting the stage for new innovations for ML in physics-based cellular simulations.

\begin{figure}[htb]
    \centering
    \includegraphics[width=\textwidth]{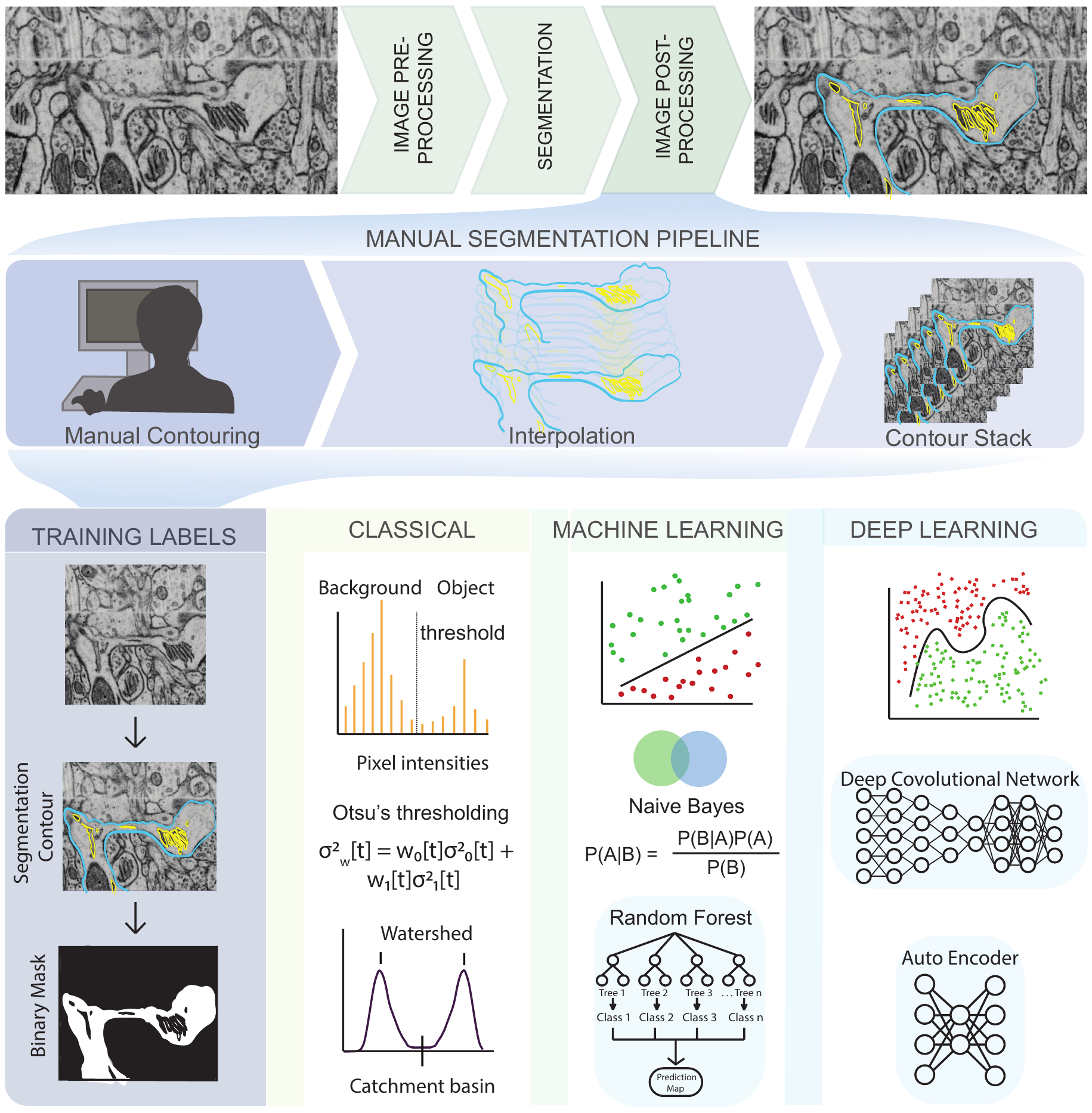}
    \caption{An illustration of the complex pipeline needed to go from imaging data to a segmented mesh, with various opportunities for emerging techniques in machine learning shown throughout the pipeline. Top row: EM images obtained from \cite{Wu2017} of dendritic spines from mouse brain tissue. Middle row: Manual tracing or contouring, interpolation, and stacking of contours is extremely time consuming, prone to error, and relies of human judgement. Bottom row: On the other hand, development of training labels and different learning techniques can reduce both time and error, bridging the gap between biological data and simulations. The list of techniques described is representative only, and not exhaustive. }
    \label{fig:MLSegmentation}
\end{figure}



\section{Sources of error in imaging modalities}

 
Images generated by the various microscopy modalities must undergo pre-processing to correct for errors such as uneven illumination or background noise \cite{van2015astra, lidke2012advances}. 
The choice of suitable algorithms for error correction depends on multiple factors, some of which are listed here -- the length scale of the experiment being conducted, scalability and reproducibility of the experiment, optical resolution of the microscope, sensitivity of the detector, specificity of the staining procedure, imaging mode (2D, 3D, 3D time-series), imaging modality (fluorescence, EM, ET etc.,) and other imaging artifacts like electronic noise, lens astigmatism, mechanical tilting/vibration, sample temperature, and discontinuous staining \cite{moen2019deep, van2015astra, lidke2012advances, koster1992automated}. 
These sources of error are an important consideration for model implementation further downstream \cite{vasan2019dlite}.


Electron tomography (ET) remains one of the most popular methods of cell imaging for modeling purposes \cite{mazel2009stochastic, west20113d, noske2008expedited}, as it retains the highest resolution of all the 3D cell imaging techniques \cite{lidke2012advances} by reconstructing a 3D object from a series of 2D images collected at different tilt angles \cite{perkins1997electron}.
However, images from ET also have a low signal to noise ratio (SNR) and have anisotropic resolution (for example, 1 nm resolution in x, y and 10 nm resolution in z) \cite{van2015astra}. 
This is partly because biological samples can withstand only a limited dose of electron beam radiation (SNR is proportional to the square root of the electron beam current) before the specimen is damaged \cite{baker2010radiation}. 
Other sources of error such as misalignment of projections and missing wedges from an incomplete tilt angular range can significantly affect the quality of the reconstruction. 
To work with data such as these, image processing steps are required for high resolution 3D reconstruction \cite{van2015astra, phan20173d}. 
Commonly used software packages for image processing such as IMOD \cite{kremer1996computer} and TomoJ\cite{messaoudii2007tomoj} use reconstruction algorithms such as Weighted Backprojection (WBP) and Simultaneous Iterative Reconstruction Technique (SIRT). 
While these have been very effective at reconstruction, sources of error can still accumulate, leading to further manual adjustment \cite{leary2013compressed}.


\section{Applications of ML for the segmentation and reconstruction of cellular structures}

Given a noisy 3D reconstruction, how can we segment cellular structures of interest? 
One approach is to employ manual segmentation tools applied to 3D tomograms such as XVOXTRACE \cite{perkins1997electron, yin2016proteolipid}, and more generally, manual contouring, interpolation, and contour stacking (Fig. 1). 
The advantage of such methods is that the human eye performs exceptionally well at detecting objects in an image \cite{moen2019deep, le2003classification}.
Consequently, semi-manual and manual segmentation are widely adopted, favoring accuracy over efficiency.
However, such methods can be extremely tedious and not always reproducible. 
Alternatively, numerous semi-automatic segmentation algorithms such as interpolation, watershed, thresholding, and clustering are available as plugins in software packages like IMOD \cite{kremer1996computer} and ImageJ \cite{abramoff2004image} (Fig. 1, classical). 
However, the accuracy of open platform algorithms is debatable \cite{jerman2016enhancement} because of two main reasons -- \textit{(i)} Even with a `perfect' ET reconstruction (no tilt misalignment, no missing wedge, no noise),  the application of filtering algorithms like Gaussian blur or non-linear anisotropic diffusion (NAD) \cite{frangakis2001noise} can cause artefacts that lead to misclassifications, rendering the image unsuitable for downstream quantitative simulations and analysis. 
\textit{(ii)} Segmentation workflows are often designed for a specific structure and/or imaging modality, limiting their generalizability and applicability.

Annual cell segmentation challenges are evidence of the demand for automatic segmentation \cite{arganda2015crowdsourcing, mavska2014benchmark}, with many of its past winners responding with ML-based programs \cite{sommer2011ilastik, ronneberger2015u}. 
Training labels for ML techniques requires a relatively small percentage (as small as 10\%) of manually segmented labels, allowing for very large data sets to be processed significantly faster than previously implemented semi-automatic segmentation methods. 
The most successful teams utilized ML techniques such as random forest classifiers, support vector machines, or a combination of these to get segmentations comparable or often even better than their human counterparts \cite{sommer2011ilastik, arganda2015crowdsourcing, ronneberger2015u, mavska2014benchmark} (Fig. 1, machine learning). 
These techniques function by imputing several image features such as noise reduction, and texture and edge detection filters \cite{arganda2017trainable}. 
These filters are then used to train a classification algorithm in an interactive manner, achieving better classification accuracy at the cost of increased training time compared to the direct application of a filter. However, because the algorithm is interactive, it still requires manual input and both the training time and accuracy can depend on the user. 

More recently, deep learning-based ML algorithms (Fig. 1, deep learning), and more specifically, convolutional neural networks (CNNs) have surged in popularity due to the success of AlexNet in the ImageNet classification challenge \cite{krizhevsky2012imagenet}. 
CNNs are complex learnable non-linear functions that do not require the imputation of data-specific features.
Indeed, CNNs learn the feature mapping directly from the image. 
The U-Net convolutional neural network architecture \cite{ronneberger2015u} further generalized deep learning, winning the ISBI neuronal structure segmentation challenge in 2015 with a quicker speed and with fewer training images.
It functions by using the feature mapping imputed by a CNN to map the classification vector back into a segmented image.
Such is the achievement of the U-Net that its variants are now the state-of-the-art in tasks like calling genetic variation from gene-sequencing data \cite{poplin2018universal}, brain tumor detection \cite{dong2017automatic} and segmentation of medical image datasets \cite{weng2019unet}. 
However, such deep learning based methods have their own challenges. They require both quality and quantity of annotated training data, significant amount of training time, graphics processing unit computing, and can generalize poorly to a different dataset. 

Both the difficulty and cost of generating annotated training data increases exponentially when dealing with Volumetric (3D) images compared with 2D, which are the desired inputs for biophysical simulations. 
Since the U-Net is a 2D architecture \cite{ronneberger2015u}, it cannot be applied directly to 3D images without modifications. 
To this end, 3D U-net used sparsely annotated 2D slices to generate volumetric segmentations of brain tumors \cite{cciccek20163d}. 
Similarly, VoxRestNet \cite{chen2018voxresnet} introduced residual learning using ResNet \cite{he2016identity}, a deep residual network capable of training hundreds to thousands of layers without a performance drop, to a voxelwise representation of 3D magnetic resonance (MR) images of the brain, paving the way for scalable 3D segmentation.

Excitingly, such algorithms are being made openly accessible and easy-to-use. 
For example, iLastik \cite{sommer2011ilastik, berg2019ilastik} and Trainable Weka Segmentation \cite{arganda2017trainable} are both available as plugins in software packages like ImageJ. 
These tools provide an interactive platform for segmentation, employing supervised classification techniques like random forests as well as unsupervised clustering such as K-means \cite{arganda2017trainable}. 
Similarly, deep learning tools such as DeepCell \cite{van2016deep} and U-Net \cite{falk2019u, ronneberger2015u} are also available in various bioimage software packages.
Other stand-alone tools like the Allen Cell Structure Segmenter provide a lookup table of 20 segmentation workflows that feed into an iterative deep learning model \cite{chen2018allen}. 
Cloud compute based segmentation plugins like CDeep3M \cite{haberl2018cdeep3m} leverage Amazon Web Services (AWS) images to provide an efficient and compute-scalable tool for both 2D and 3D biomedical images. 

Generating well-organized and annotated training data continues to be the major challenge for most ML segmentation methods. 
Crowdsourced annotation tools like Amazon's Mechanical Turk can be useful in this context, but are still limited by the difficulty of training naive users on tracing specific structural images. 
Alternatively, many ML algorithms leverage transfer learning approaches using pre-trained networks such as VGG-net \cite{bruna2015super, johnson2016perceptual,simonyan2014very}, AlexNet \cite{krizhevsky2012imagenet} and GoogleNet \cite{szegedy2015going}.
In fact, popular semantic segmentation and clustering networks like Fully Convolutional Networks (FCN) \cite{long2015fully} and DECAF \cite{donahue2014decaf} are themselves implemented using transfer learning approaches. Such transfer learning can also be used to generalize models trained on biological data to a different cell type or experimental condition, significantly reducing the time for training and accompanying computing resources required. More recently, label-free approaches employing a U-net variant have been applied to predict cellular structure from unlabeled brightfield images \cite{ounkomol2018label, christiansen2018silico}. 
These methods can serve as a platform for building low cost, scalable, and efficient segmentation of 3D cellular structure.

\section{Applications of ML for the generation of synthetic cellular structures}

There are two main aspects involved in the development of comprehensive biophysical models -- (1) what is the process being modeled? and (2) what is the geometry in which this process is being modeled? 
Answers to the first question are based on experimental observations and specific biology. 
Answering the latter is significantly more challenging because of the difficulties in -- \textit{(i)} obtaining accurate segmentations, \textit{(ii)} discovering new structure from experiments, and \textit{(iii)} simultaneously visualizing multiple structures. 
The use of synthetically generated geometries, which can probe different arrangements of organelles within cells could be relevant for generating biologically relevant hypotheses.  


A subset of ML models, called \textit{generative} models, deal with the task of 
generating new synthetic but realistic images that match the training set distribution.
For our purposes, such methods are relevant in the context of generating \textit{(i)} noise-free images, \textit{(ii)} images representative of a different cell type, structure, or time-point, and \textit{(iii)} unlikely images that represent the most unique shapes of the structure being imaged.
For example, by capturing the unlikely and likely shapes in our dataset, we could generate sequences of synthetic images that transition from one shape to the next. 
These synthetic images can be used in biophysical simulations to generate biologically relevant hypotheses. 




In recent years, there has been rapid progress in applying deep generative models to natural images, text, and even medical images. Popular classes of deep generative models like Variational Autoencoders \cite{kingma2013auto}, Generative Adversarial Networks \cite{goodfellow2014generative}, and Autoregressive models such as PixelRNN \cite{oord2016pixel} and PixelCNN \cite{van2016conditional} have achieved state of the art performance on popular image datasets such as MNIST \cite{deng2012mnist}, CIFAR \cite{krizhevsky2009learning} and ImageNet \cite{deng2009imagenet}. 
Each class of models has numerous modified implementations. For example, GANs alone include models like deep convolutional GAN (DCGAN) \cite{radford2015unsupervised}, conditional GAN (cGAN) \cite{antipov2017face}, StackGAN \cite{zhang2017stackgan}, InfoGAN \cite{chen2016infogan} and Wasserstein GAN \cite{arjovsky2017wasserstein} to name a few. 
Each model has its own distinct set of advantages and disadvantages. 
GANs can produce photo-realistic images at the cost of tricky training and no dimensionality reduction. 
VAEs allow for both generation and inference, but their naive implementation results in less photo-realistic generative examples. 
Autoregressive models obtain the best log-likelihoods at the cost of poor dimensionality reduction. 
Importantly, all of these models are unsupervised, implying that they are not limited by manual annotation that is otherwise a common challenge to supervised learning approaches.




In cell biology, much of the work in building generative models of cellular structures has been associated with the open source CellOrganizer \cite{johnson2015joint, johnson2015automated, shariff2010automated, rohde2008deformation, shariff2011automated, peng2011image, zhao2007automated}, which uses a Gaussian Mixture Model given reference frames like the cell and nuclear shape in order to predict organelle shape distribution. 
These models also have the option to be parametric (parameters such as number of objects), which reduces the complexity of the learning task, the training time and GPU computing resources required, while also allowing for exploration and analysis of the parameters and their effect on the spatial organization of cells. 
Aside from CellOrganizer, other recent efforts have begun to leverage deep generative models in cell biology. We now have models that can predict structure localization given cell and nuclear shape \cite{johnson2017generative}, extract functional relationships between fluorescently tagged proteins structures in cell images \cite{osokin2017gans}, learn cell features from cell morphological profiling experiments \cite{caicedo2018weakly}, and interpret gene expression levels from single-cell RNA sequencing data \cite{lopez2018deep, ding2018interpretable}. 




The challenge going forward will be how best to use generative modeling given the data in hand. This will depend on the question we want to ask of the data. For example, if we are modeling processes associated with cell and nuclear shape, spherical harmonics based generative models might be more appropriate than deep learning based methods \cite{ruan2018evaluation}. 
If we are interested in inpainting a missing wedge from a tomogram using a generative model, then GANs might be more appropriate \cite{ding2019joint}. Generated images can also be used as a source of training data for segmentation and classification tasks \cite{yang2019learning}.
Taken together, these choices will help develop efficient end-to-end pipelines for segmentation and shape generation, and provide a platform for running biophysical simulations.
Already, CellOrganizer can export spatial instances to cell simulation engines such as MCell \cite{stiles2001monte} and VirtualCell \cite{loew2001virtual}, allowing us to simulate chemical reactions in different spatial compartments. 
Similar pipelines for deep generative models will need to be implemented in order to fully realize their downstream interpretations.

\section{Applications of ML for meshing, simulation, and data analysis}

\begin{figure}[htb]
    \centering
    \includegraphics[width=\textwidth]{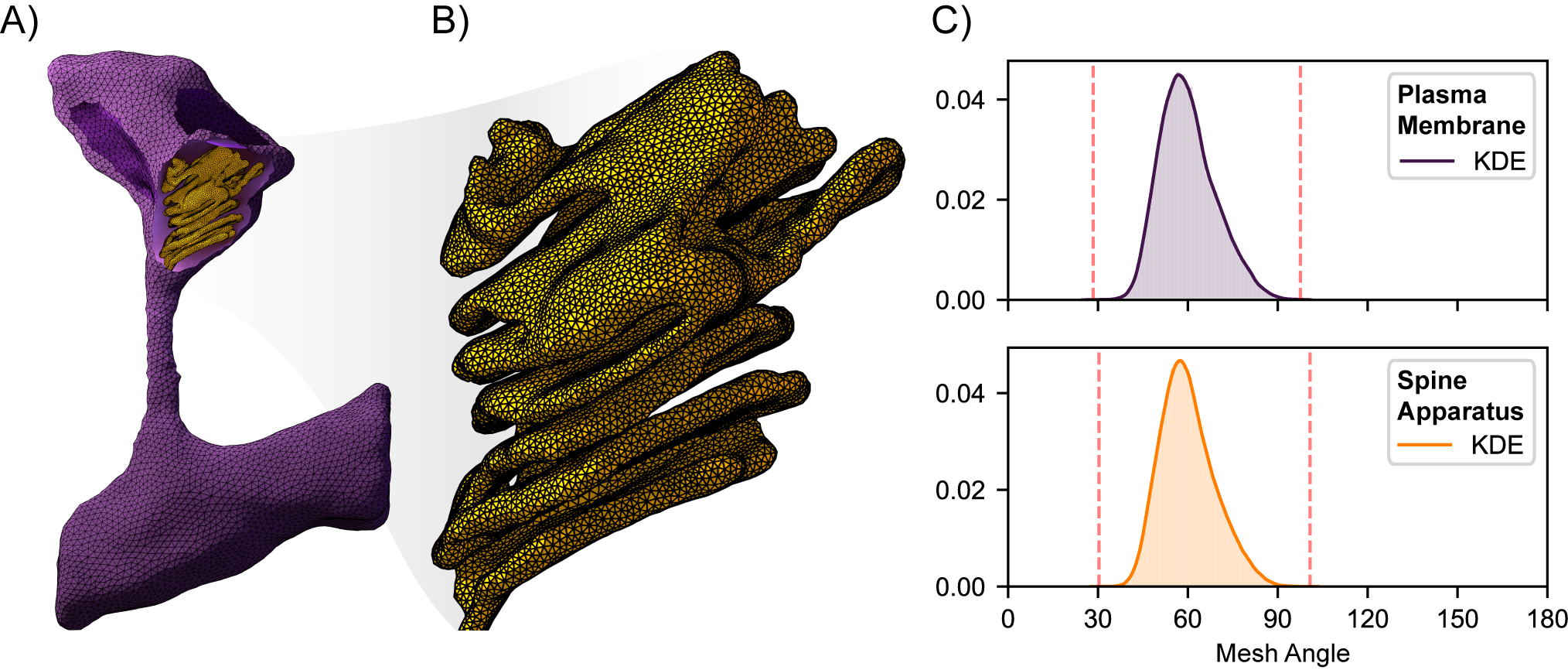}
    \caption{An illustration of complexity, size, quality, and local resolution of meshes typically needed for realistic simulation of biophysical systems.
    Meshes are generated using \texttt{GAMer 2} \cite{LLBA19a,LLMA19a}.
    A) Example surface mesh of a dendritic spine with geometry informed by electron micrographs from \cite{Wu2017}.
    The plasma membrane is shown in purple with the post synaptic density rendered in dark purple. The spine apparatus, a specialized form of the endoplasmic reticulum is shown in yellow.
    B) A zoomed in view of the spine apparatus. 
    Note that the mesh density is much higher in order to represent the fine structural details.
    C) Binned histogram distributions of mesh angles for both the plasma membrane and spine apparatus. 
    The colored smooth lines are the result of a kernel density estimate.
    Dotted red lines correspond to the minimum and maximum angle values in each mesh.
    Both meshes are high quality with few high aspect ratio triangles (i.e., those deviating most from equilateral).
    }
    \label{fig:GAMer}
\end{figure}

ML is commonly applied to mesh segmentation and classification; examples include PointNet~\cite{qi2017pointnet} (segments and classifies a point cloud), and MeshCNN~\cite{hanocka2019meshcnn} (segments and classifies edges in a mesh).
However, although the term \textit{machine learning} was not traditionally used to describe meshing techniques, in fact algorithms for mesh generation (cf.~\cite{LMAM18a}), mesh improvement (such as mesh smoothing~\cite{BaSm97}), and mesh refinement~\cite{LiJo95,Maub95,Bey95,AMP97} all fundamentally involve local (cf.~\cite{GYH12}) and/or global (cf.~\cite{ChHo10a}) optimization of an \textit{objective function} (see Fig. 2).
Mesh point locations, and/or edge/face connectivity decisions are viewed as parameters that are determined (or \textit{learned}) as part of an iterative algorithm that extremizes a local or global objective function (usually involving constraints as well) in an effort to generate, improve, or refine a given mesh.  In addition, \textit{adaptive numerical methods} for simulation of physical systems involving the solution of ordinary (ODE) and partial (PDE) differential equations are again an early example of the application of ML techniques in computational science, long before the terminology was widely used.
A classic reference from the 1970's in the context of adaptive finite element methods is~\cite{BaRh78a,BaRh78b}; all modern approaches to adaptive numerical methods for  ODE and PDE systems continue to follow the same general framework outlined in that work: \textit{(i)} Solve the ODE/PDE on the current mesh; \textit{(ii)} Estimate the error using \emph{a posteriori} indicators; \textit{(iii)} Refine the mesh using provably non-degenerate local refinement with closure; \textit{(iv)} Go back to step \textit{(i)} and repeat the iteration until a target quality measure is obtained (a standard approach is to approximately minimize a \emph{global error function}, through the use of local error estimates).
These types of adaptive algorithms are effectively \textit{machine learning} the best possible choice (highest accuracy with least cost) of mesh and corresponding numerical discretization for the target ODE/PDE system.
Recent work in the area is now moving toward a more explicit and sophisticated use of modern ML techniques (cf.~\cite{fmats2019,Manevitz2005NeuralNT}). ML can further assist in simulation and data analysis further downstream. Specifically, it can accelerate \textit{(i)} parameter estimation, \textit{(ii)} uncertainty quantification, and \textit{(iii)} dimensionality reduction, three of the most common post-processing tasks from a biophysical simulation. Incorporating specific biophysical model information like stress-strain relationships \cite{mendizabal2019simulation} or statistical molecular dynamic states \cite{noe2019boltzmann} into ML algorithms can also reduce the computational time for numerical solvers. Finally, more standard ML approaches like clustering and dimensionality reduction can assist in both visualization and interpretation of simulation results.

\section{Perspectives and Future Directions}

In this perspective, we have discussed three key aspects of a pipeline for realistic cellular simulations: \textit{(i)} Reconstruction and segmentation of cellular structure; \textit{(ii)} Generation of cellular structure; and \textit{(iii)} Mesh generation, refinement and simulation. While these were discussed separately, neural networks like Pixel2Mesh demonstrate the feasibility of end-to-end pipelines from a single black box \cite{wang2018pixel2mesh}. Of course, black boxes are not interpretable, and recent ML frameworks like SAUCIE have begun to use regularizations to enforce mechanistic intepretability in the hidden layers of an autoencoder neural network \cite{amodio2019exploring}. We anticipate that future endeavours will implement a fully interpretable and end-to-end pipeline for biophysical simulations.

\section*{Conflict of Interest Statement}

The authors declare that the research was conducted in the absence of any commercial or financial relationships that could be construed as a potential conflict of interest.

\section*{Author Contributions}

All co-authors contributed to the writing of the manuscript,
and also provided area-specific expertise.

\noindent
\textit{R. Vasan}: Provided expertise on generation and simulation.

\noindent
\textit{M. Rowan}: Provided expertise on imaging, segmentation, and reconstruction.

\noindent
\textit{C.T. Lee}: Provided expertise on reconstruction, meshing and simulation.

\noindent
\textit{G.R. Johnson}: Provided expertise on imaging, segmentation, reconstruction and generation.

\noindent
\textit{P. Rangamani}: Provided expertise on imaging, segmentation, reconstruction, and simulation.

\noindent
\textit{M. Holst}: Provided expertise on meshing, simulation, and analysis.

\section*{Funding}

R.V., M.R., C.T.L, and P.R. are supported in part by an AFOSR MURI award FA9550-18-1-0051 and ONR fund ONR N00014-17-1-2628.
C.T.L. also acknowledges support from a Hartwell Foundation Postdoctoral Fellowship.
M.J.H. was supported in part by NSF Awards 1934411 and 1630366.

\section*{Acknowledgments}

We would like to thank Prof. Pietro De Camilli and coworkers for sharing their datasets from Wu
et al. \cite{Wu2017}. We also thank Dr. Matthias Haberl, Mr. Evan Campbell, Profs. Brenda Bloodgood and
Mark Ellisman for helpful discussion and suggestions. 
GJ thanks the Allen Institute for Cell Science founder, Paul G. Allen, for his vision, encouragement and support.



\bibliographystyle{frontiersinHLTH&FPHY} 
\bibliography{collection}


\clearpage

\newpage

\end{document}


\onecolumn
\firstpage{1}

\title[Supplementary Material]{{\helveticaitalic{Supplementary Material}}}

\maketitle

\section{Supplementary Data}

Supplementary Material should be uploaded separately on submission. Please include any supplementary data, figures and/or tables. All supplementary files are deposited to FigShare for permanent storage and receive a DOI.

Supplementary material is not typeset so please ensure that all information is clearly presented, the appropriate caption is included in the file and not in the manuscript, and that the style conforms to the rest of the article. To avoid discrepancies between the published article and the supplementary material, please do not add the title, author list, affiliations or correspondence in the supplementary files.

\section{Supplementary Tables and Figures}

For more information on Supplementary Material and for details on the different file types accepted, please see \href{http://home.frontiersin.org/about/author-guidelines#SupplementaryMaterial}{the Supplementary Material section} of the Author Guidelines.

Figures, tables, and images will be published under a Creative Commons CC-BY licence and permission must be obtained for use of copyrighted material from other sources (including re-published/adapted/modified/partial figures and images from the internet). It is the responsibility of the authors to acquire the licenses, to follow any citation instructions requested by third-party rights holders, and cover any supplementary charges.


\subsection{Figures}


\begin{figure}[htbp]
\begin{center}
\includegraphics[width=9cm]{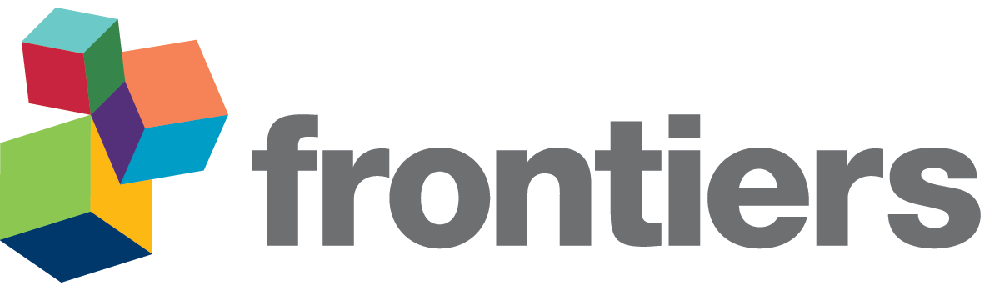}
\end{center}
\caption{ Enter the caption for your figure here.  Repeat as  necessary for each of your figures}\label{fig:1}
\end{figure}

\begin{figure}[htbp]
\begin{center}
\includegraphics[width=10cm]{logos}
\end{center}
\caption{This is a figure with sub figures, \textbf{(A)} is one logo, \textbf{(B)} is a different logo.}\label{fig:2}
\end{figure}


\bibliographystyle{frontiersinHLTH&FPHY} 